\documentclass[review]{elsarticle}
\usepackage{lineno,hyperref}
\usepackage{times}
\usepackage{epsfig}
\usepackage{graphicx}
\usepackage{dcolumn}
\usepackage{bbold}
\usepackage{epsfig}
\usepackage{epsf}
\usepackage{lscape}
\usepackage{tabularx}
\usepackage{color}

\usepackage{natbib}

\modulolinenumbers[1]

\begin{document}
\begin{frontmatter}
\title{Relativistic equation-of-motion coupled-cluster method for the electron attachment problem}
\author{Himadri Pathak,$^{1,\ast}$ Sudip Sasmal,$^{1,\dag}$ Malaya K. Nayak,$^2$ Nayana Vaval,$^1$ and Sourav Pal$^{3}$}
\address{$^1$Electronic Structure Theory Group,
Physical Chemistry Division,
CSIR-National Chemical Laboratory,
Pune, 411\,008, India}
\address{$^2$Bhabha Atomic Research Centre, Trombay, Mumbai-400\,085, India}
\address{$^3$Department of Chemistry, Indian Institute of Technology Bombay, Powai, Mumbai 400\,076, India}
\fntext[Himadri Pathak]{$^\ast$hmdrpthk@gmail.com}
\fntext[Sudip Sasmal]{$^\dag$sudipsasmal.chem@gmail.com}


\begin{abstract}
The article considers the successful implementation of relativistic equation-of-motion coupled cluster method for the
electron attachment problem (EA-EOMCC) at the level of single- and double- excitation approximation.
The implemented relativistic EA-EOMCC method is employed to calculate ionization potential
values of alkali metal atoms (Li, Na, K, Rb, Cs, Fr) and the vertical electron affinity values of 
LiX (X=H, F, Cl, Br), NaY (Y=H, F, Cl) starting from their closed-shell configuration.
Both four-component and exact two component calculations are done for all the opted systems. 
Further, we have shown the effect of spin-orbit interaction considering the atomic systems. 
The results of our atomic calculations are compared with the values from the NIST database  
and the results are found to be very accurate ($<1\%$).
\end{abstract}


\end{frontmatter}

\clearpage
\newpage
\section{Introduction}
A considerable growing interest is noticed in recent years in the study of negative ions as negative ions have significance
in many areas of physics like in astrophysics, plasma physics and surface physics \cite{0,1,2,3}. The electron affinity
(EA) is an important quantity of these ions. The precise measurement of EA of atomic or molecular systems is always a challenge
as the resulting negative ion is difficult to handle. Despite of the complexity in the measurement, there have been significant advances in
the experimental techniques like laser photodetachment electron spectroscopy (LPES), laser photodetachment
threshold spectroscopy (LPTS), accelerator mass spectroscopy (AMS) and photodetachment microscopy, et cetera that
are capable of precise measurements of EA of an atomic system \cite{3a,3b,3c,3d}. However, the situation is inappreciative
in achieving such an extent of accuracy in the
molecular systems due to the possibility of structural change on attachment of an extra electron. Therefore, it is
an outstanding challenge for the computational physicists to complement these atomic measurements as well as
for new predictions for the future purpose.\par
The computational prediction of EA is difficult due to the absence of long-range Coulomb field
outside of a neutral precursor. Therefore, an extra electron is solely bound through correlation with
other electrons \cite{3e,4}. Moreover, most of the theoretical calculations are based on the
quantum chemical basis set methods. Thus, the finite size of the basis and unbalanced treatment
of electron correlation in the atomic or molecular system and in the resulting ion are the sources of error \cite{5,6}.
The attachment and detachment of an electron to a neutral species involves
different forces. The attached extra electron to the neutral atom polarizes the
electronic shell of the atom. As a result a dipolar electric field is generated
which binds the extra electron with the other electrons. The charge distribution
of the electron cloud, particularly the electron-electron correlation effects
decides the stability of the negative ion. These interactions do not play much
role in most of the neutral atoms as well as in positive ions where direct electrostatic
force is the dominant factor for the stability of the neutral atom or
the positive ion. On the other hand, these effects dominate in the negative ions.
Therefore, the calculations of EA values of both atomic and molecular systems
are challenging and is a real test for the performance of a many-body method.
It is an established fact that not only the electron correlation but also the effect of relativity play a definite role in accurate description 
of the eigenstates of heavy atomic and molecular systems \cite{7a}. 
It is, therefore, in such a case a highly correlated many-body method, capable of simultaneous treatment of
relativity and electron correlation is required due to the intricate coupling between these two effects \cite{8,9,9a}. \par

The relativity has a greater role towards the core orbitals and practically important for all the elements.
The effects of relativity are incorporated in the electronic structure calculations by the choice of the Hamiltonian. 
The consideration of Dirac-Coulomb-Breit Hamiltonian without the quantum electrodynamics effects (QED) is sufficient
for most of the relativistic electronic structure 
calculations using four-component wavefunction. However, in actual practice the Dirac-Coulomb Hamiltonain is most commonly used where 
two-body Coulomb interaction operator is added to the Dirac Hamiltonian ($\hat H_D$).
Although, the form of the Coulomb operator is same 
as in the non-relativistic theory, however, the physical content is different as it takes care 
of the spin-same orbit interaction. 
This type of truncation in the two-body interaction does not effect much for most of the chemical purposes \cite{visser1992relativistic}.
However, for very accurate studies of molecular spectra including fine structure, the inclusion of spin-other-orbit 
interaction and spin-spin interaction are required which can be done with the full inclusion of the Breit part of the two-body interaction.
The relativistic calculations using four-component wave function are very expensive from the computational perspective.
A lot of effort has been made to simplify the equations. The calculation of the small component of the wave function 
is the most challenging part of the computation.
If a basis set is expressed in terms of contracted Gaussian functions,
then the number of required primitive Gaussian functions for the small component 
is about twice the number of the large component with the imposition of the kinetic balance condition. On the other hand, the small 
component has a very minor contribution in the calculated values; therefore, it makes sense to look for an approximation.
There are a number of Hamiltonians in between the scalar non-relativistic and four-component relativistic
ones. However, the inclusion of the spin-orbit interaction requires at least a two-component description, though it will
essentially increase the computational cost due to the appearance of complex algebra in place of real algebra. 
The electron correlation methods in the no-pair approximation require the transformation of the matrices from the atomic 
orbital (AO) basis to the molecular orbital (MO) basis. The spin coordinates of the electrons can be represented in terms of 
quaternion algebra in the four-index transformation step which helps to go from complex four-component to a two-component
quaternion form. Therefore, the MO coefficients become  quaternion and can be represented in terms of real matrices \cite{saue, visscher_2002}. 

The generation of a two-component Hamiltonian from the parent four-component Hamiltonian is the most preferred choice for the purpose
which includes the spin-orbit interaction with a lesser cost as compared to the four-component Hamiltonian. 
The central idea behind the generation of a two-component Hamiltonian is that it should reproduce the positive-energy spectrum of the parent
Hamiltonian. Foldy and Wouthuysen proposed an idea to decouple the large and small component
by a unitary transformation of the four-component Hamiltonian. 
Another well known approach is the elimination of the small component from the wavefunction. 
However, these two approaches can be shown to be equivalent \cite{IOTC}.
The exact two-component approach (X2C) in the two-component framework is one such approach to reduce the computational scaling which uses
elimination of the small component from the parent four-component Hamiltonian.  
The detailed description of the X2C approach including various other two-component methods can be found in
Ref. \cite{7a, saue2011relativistic, saue2012relativistic}. 
%
\par

Over the years, the equation-of-motion coupled-cluster (EOMCC) method gained popularity among correlation methods for the treatment of electron correlation due to its 
simplicity and elegance. 
The idea of EOMCC \cite{14,16,17,19,20,23,25,piecuchwa1,piecuchwa2} is conceptually very simple and it is operationally a two step process:
(i) solution of coupled cluster problem with the $N$ electron closed-shell determinant as reference and
(ii) construction and diagonalization of the effective Hamiltonian matrix for the
Fock-space sector of interest in the configuration space.
It takes into account of both the dynamic and non dynamic part
of the electron correlation. The exponential structure of the coupled cluster operator takes care of the dynamic part of the electron correlation 
and non dynamic part is included by means of diagonalization of the effective Hamiltonian
matrix in the configurational space.
The diagonalization of effective Hamiltonian, by and large is associated with the multi-reference theories,
whereas EOMCC works within a single reference description to tackle the complex multi-configurational wavefunction.
Further, the relaxation effect, which has an important
role in proper description of the eigenstates is also taken care.  
The multiple roots can be addressed in a single calculation and each of the states are treated with equal weightage.
The EOMCC method behaves properly at the non-interacting limit but not rigorously extensive (only for the core-core and core-valence interactions)
due to the linear structure of the EOM operator \cite{26}.
The EOMCC is in close kinship with the coupled cluster linear response theory (CCLRT) \cite{26b,26e} and symmetry adapted cluster
expansion configuration interaction (SAC-CI) method \cite{26f,26g}.
It is worth to note that the transition energy calculated using CCLRT is identical with the EOMCC method for the one valence problem
but the transition moments is identical only when it is represented as a energy derivative in EOMCC framework.
Chaudhuri {\it et al} \cite{rajat,rajat2} applied relativistic CCLRT for the ionization problem of atomic systems with spherical
implementation.
Besides these two methods, effective Hamiltonian variant of
Fock space multi-reference (FSMRCC) theory \cite{26aa,26ae,26af,26ag,26aj,26ak} always comes in the discussion on EOMCC as these two methods
produce identical results for the one valence problem. The amplitudes of all the lower sector including the sector of interest are involved 
in the FSMRCC theory. On the other hand, EOMCC deals with the amplitudes of the (0,0) sector and the sector of interest.
Therefore, both the approaches are eventually produce the same result for the one electron attachment or detachment problem.
The EOMCC is free from the problem of intruder due to its CI
(configuration interaction method) like structure, which is associated with the effective Hamiltonian variant of the FSMRCC theory. 
There are ways in the FSMRCC framework to handle the problem of intruder such as the eigenvalue independent partitioning technique of Mukherjee
(EIP-FSMRCC) \cite{26,EIP} and the intermediate Hamiltonian variant of the FSMRCC (IH-FSMRCC) theory \cite{IH1,IH2,IH3}. \par  
Recently, Blundell implemented relativistic EOMCC method for the electron affinity 
problem and applied to calculate fine-structure splittings in high-lying states of rubidium atom \cite{27}.
The implemented version of Blundell is applicable only for the purpose of atomic calculations as they have used the spherical
implementation which allows the separation of radial and angular parts.
Therefore, the evaluation of radial integrals is only required and the angular part will add up to it as a multiplier.
The radial integrals can be evaluated numerically.
Such a separation is not possible in molecular systems due to the absence of spherical symmetry.
In our implemented version, we have used the one- and two- body matrix elements, which are evaluated in the
Cartesian coordinate system. 
The Cartesian coordinate system does not allow one to exploit the spherical symmetry to separate the matrix elements
into radial and angular parts.
Furthermore, the anti-symmetrized two-body matrix elements are used in this coordinate system
calculations, which is not possible in the spherical implementations as angular factor will 
be different for the direct and exchange part of the two body matrix element. 
Thus, our implemented version is a general one, applicable to both atoms as well as molecules starting from their closed-shell
reference state configuration.
It should be noted that the spherical implementation is much more complex than that of the molecular calculations, but
it is favorable from the computational point of view as it requires only the solution of radial integrals.
Therefore, atomic calculations are computationally easy, which allows to correlate
more number of electrons and amenable to use huge basis for the correlation
calculation to achieve a better accuracy. \par
In our recent work,
the performance of the fully four-component EOMCC has already been established for both atomic and molecular systems
for the single ionization and double ionization problem  \cite {10,11,12}.
Therefore, in this article, we focus on the implementation of relativistic EOMCC method for the electron
affinity problem applicable to both atomic and molecular systems.
The implemented EA-EOMCC method is employed to calculate ionization potential of open-shell atomic 
systems starting from their singly positive closed-shell configuration. Further, the vertical EA values of molecular systems
are also calculated.
Both four-component and exact two component (X2C) calculations are done 
for all the considered systems. The effect of spin-orbit interaction is shown for the atomic systems in the EOMCC framework. \par
The manuscript is organized as follows.
The EOMCC theory in regard to the electron attachment problem is briefly described in Sec. \ref{sec2}
and the computational details of our calculations are presented in Sec. \ref{sec3}.
We have discussed our results in Sec. \ref{sec4} and finally made concluding 
remarks in Sec. \ref{sec5}.
We are consistent with the atomic unit if not stated explicitly. \par
\section{Theory}\label{sec2}
\begin{figure}[h]
\centering
\begin{center}
\includegraphics[height=2.0cm, width=7.2cm]{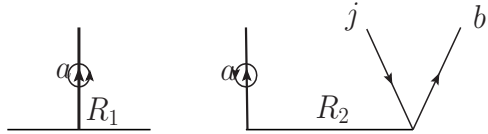}
\caption { Diagrammatic representation of $R_1$ and $R_2$ operator.}
\label{fig1}
\end{center}  
\end{figure}

In the EOMCC method the k$^{th}$ target excited state of single electron attached state
is defined as
\begin{eqnarray}
 |\Psi_k\rangle=R_k^{N+1}|\Psi_0\rangle, k=1,2,\dots
\end{eqnarray}
Here, the $R_k^{N+1}$ is a linear operator, which on acting on the single reference coupled cluster (SRCC) ground 
state wave function $|\Psi_0\rangle$, generates the $k^{th}$ excited state wave function $|\Psi_k\rangle$.
The $R_k^{N+1}$ operator takes the form in the coupled cluster single-double (CCSD) approximation as 
\begin{eqnarray}
R_{k}^{N+1} &=& R_{1} + R_{2} \nonumber  \\
            & =& \sum_{a} r^{a} a^{\dag}_a + \sum_{b<a} \sum_{j} r_{j}^{ba} a^{\dag}_a a^{\dag}_b a_j
\end{eqnarray}
The $R_1$ and $R_2$ operator are diagrammatically represented in Fig. \ref{fig1}.
The $R_1$ is a one particle (1p) creation operator and $R_2$ is a two-particle and one-hole (2p-1h)
creation operator. The circled arrow is just to represent that overall it is a one electron 
attachment process.

\begin{figure}[ht]
\centering
\begin{center}
\includegraphics[height=6.5 cm, width=10.0 cm]{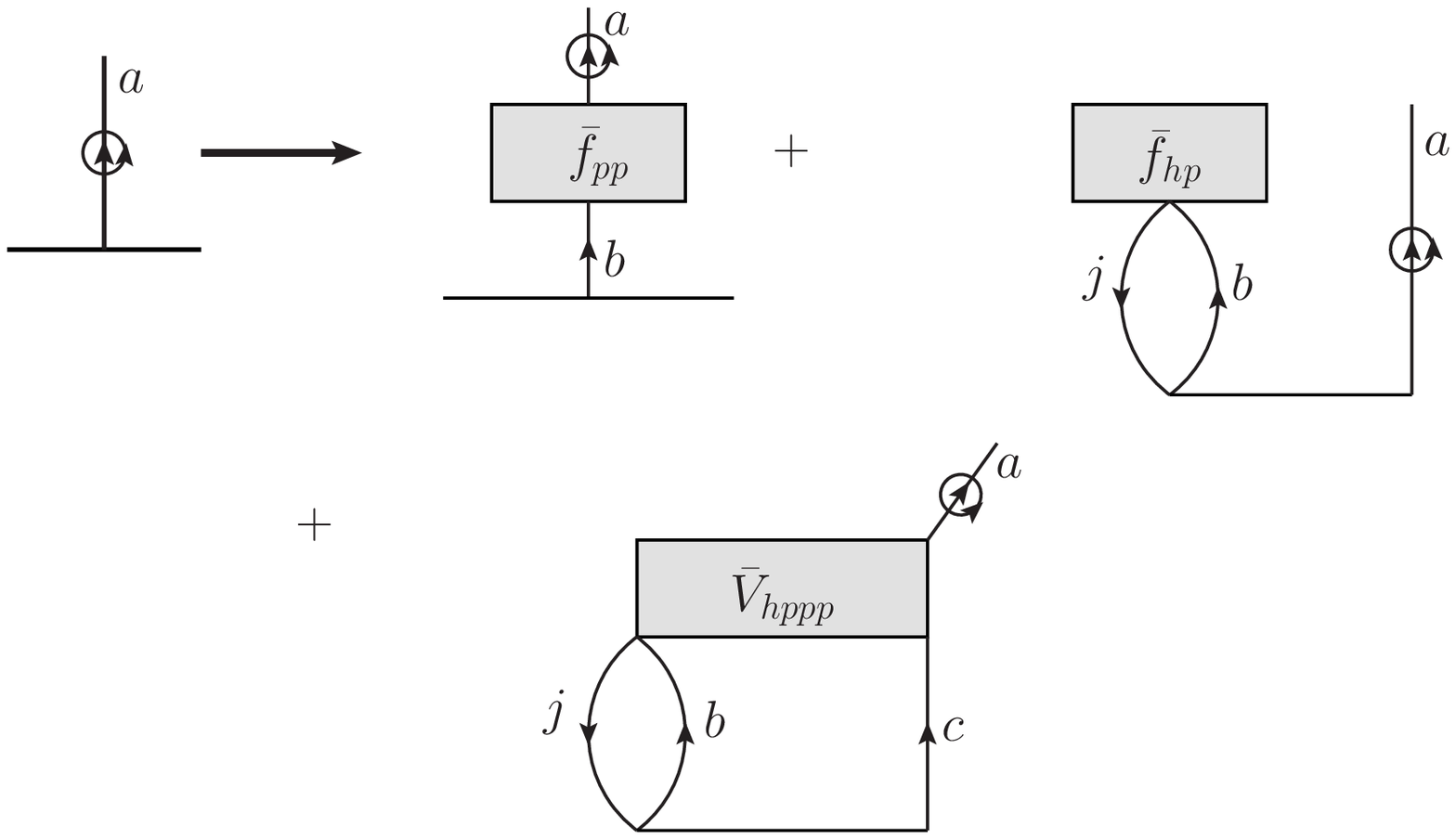}
\caption {Diagrams contributing to the 1p block.}
\label{fig2}
\end{center}  
\centering
\begin{center}
\includegraphics[height=9 cm,width=10.0 cm]{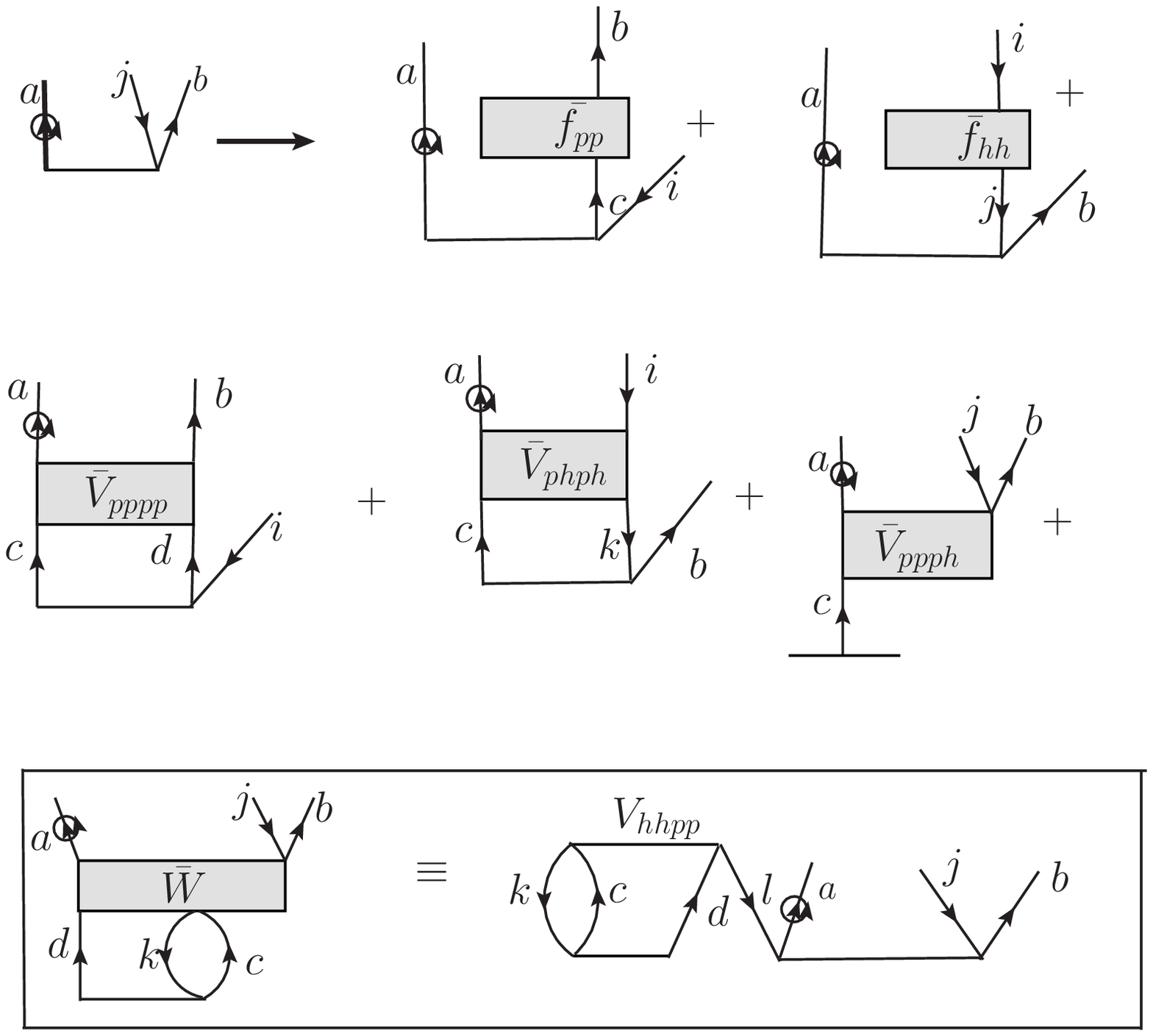}
\caption {Diagrams contributing to the 2p-1h block.}
\label{fig3}
\end{center}  
\end{figure}

The Schr\"{o}dinger equation for the ground state (k=0) is 
\begin{eqnarray}
H_{N}|\Psi_{0}\rangle=\Delta E_{0}|\Psi_{0}\rangle, 
\end{eqnarray}
The electron attached states (k=1,2,$\dots$) is written as   
\begin{eqnarray}
H_{N}R_k|\Psi_{0}\rangle=\Delta E_{k}R_k|\Psi_{0}\rangle
\end{eqnarray}
The above equation on multiplication with a non-singular operator $e^{-T}$ (where T is the coupled cluster excitation operator) in the course it is assumed 
that $R_{k}$ commute with $T$ (as strings of same quasi-particle creation operator) with some mathematical manipulation leads to
equation of motion with respect to the $R_k$ operator,
\begin{eqnarray}
[\bar{H}_N,R_k]|\Phi_{0} \rangle= \Delta E_k R_k|\Phi_{0} \rangle\,\,\, \forall k.
\end{eqnarray}
In the above equation, $\Delta E_k$ is the energy change associated with the electron attachment process
and ${\bar{H}_N = e^{-T} H e^{T}}-\langle \Phi_{0}|e^{-T}He^{T}|\Phi_{0}\rangle $ is the similarity 
transformed normal ordered effective Hamiltonian.
In our case it is the Dirac-Coulomb Hamiltonian, which is given by  
\begin{eqnarray}
{\hat H_{DC}} &=& \hat H_D+ \sum_{i>j} \frac{1}{r_{ij}} {\mathbb{1}_4} \nonumber \\
        &=&\sum_{A}\sum_{i} [c (\vec {\alpha}\cdot \vec {p})_i + (\beta -\mathbb{1_4}) c^{2} + V_{iA}] 
       + \sum_{i>j} \frac{1}{r_{ij}} {\mathbb{1}_4},
\end{eqnarray}
where $\mbox{\boldmath$\alpha$}_i$ and $\beta$ are the usual Dirac matrices, $V_{iA}$ is
the nuclear potential and $\frac{1}{r_{ij}}$ is the electron-electron repulsion potential.
The orbital energies are scaled with respect to the free electron rest mass energy ($c^2$), which 
is zero in the non-relativistic case.
We have chosen a correlated determinantal space of  $|\Phi^{a}\rangle$ and $|\Phi^{ab}_{j}\rangle$ (1p and 2p-1h)
with respect to the Dirac-Hartree-Fock determinant ($|\Phi_{0}\rangle$) to project the above equation to get the desired electron affinity
values, $ \Delta E_k$.  
\begin{equation}
{\langle \Phi^{a}[\bar{H}_N, R_k]|\Phi_{0}\rangle=\Delta E_k R^a},
\end{equation}
\begin{equation}
{\langle \Phi^{ab}_{j}[\bar {H}_N,R_k]|\Phi_{0} \rangle =\Delta E_k R^{ba}_{j}},
\end{equation}
In Figs. \ref{fig2} and \ref{fig3}, the contributing diagrams for the 1p and 2p-1h space is presented, respectively.
The evaluation of these diagrams requires the solution of the coupled-cluster amplitude equations. 
The converged $T_1$ and $T_2$ amplitudes are contracted with the two-body matrix elements of the 
Hamiltonian matrix to construct one body, two body and three body intermediate diagrams.
The intermediate diagrams are categorized into $\bar f_{pp}$, $\bar f_{hp}$,   $\bar f_{hh}$,
$\bar V_{hppp}$, $\bar V_{pppp}$,   $\bar V_{phph}$, $\bar V_{ppph}$ and $\bar W$. 
Here $\bar f$'s, $\bar V$'s and   $\bar W$ stands for one-body, two-body, and three-body
intermediates, respectively. We have followed a recursive intermediate factorization scheme
as described in Ref. \cite{30a} to evaluate these intermediate diagrams.
The factorization scheme in the construction of intermediate diagrams saves enormous
computational resources. The matrix elements corresponding to the three-body intermediate
diagram are not stored rather computed on the fly.
The programmable algebraic expression for the diagrams corresponding to 
the projection of Hamiltonian to
1p and 2p-1h excited determinantal space are 
presented in Eq. \ref{r1equation} and \ref{r2equation}, respectively.
We have used the standard notation ($\bar f$(out, in)) and ($\bar V$ (left out, right out, left in, right in))
for one-body and two-body intermediate matrix element. In Eqn \ref{r1equation} and \ref{r2equation}, 
$i, j,\dots (a, b \dots)$ stands for hole (particle) index. $\hat P$ is a permutation operator 
and any odd permutation introduces a negative sign.
The Davidson algorithm \cite{31} is implemented to get the desired set of eigenvalues and eigenvectors.
\begin{eqnarray}
\Delta E_k R^a=\sum_b \bar f_{pp}(a,b)r^b+\sum_{j,b} \bar f_{hp}(j,b)r_j^{ba}
+0.5\sum_{j,b,c} \bar V_{hppp}(j,a,b,c)r_j^{bc} \,\,\, \forall \,\,a
\label{r1equation}
\end{eqnarray}
\begin{eqnarray}
\Delta E_k R^{ba}_{j}=\hat P(ab)\sum_c \bar fpp(b,c)r_i^{ca}-\sum_j \bar f_{hh}(j,i)r_j^{ba}+
0.5\sum_{c,d} \bar V_{pppp}(a,b,c,d)r_i^{dc} \nonumber \\
-\hat P(ab)\sum_{c,k}\bar V_{phph}(a,k,c,i)r_k^{bc}
+\sum_{c}\bar V_{ppph}(a,b,c,j)r^c \nonumber \\
-0.5\sum_{k,l,c,d} V_{hhpp}(k,l,c,d)r_k^{cd}t_{lj}^{ab} \,\,\, \forall\,\,\, (i,\,\,b<a)
\label{r2equation}
\end{eqnarray}
\begin{landscape}
\begin{table}[ht]
\caption{SCF (${E_{DF}^{0}}$) and correlation energy from MBPT(2) and CCSD of alkali metal ions.}
\newcommand{\mc}[3]{\multicolumn{#1}{#2}{#3}}
\begin{center}
\begin{tabular}{lrccrcc}
\hline
\hline
 & \mc{3}{c}{X2C} & \mc{3}{c}{4C}\\
\cline{2-4} \cline{5-7}\\
Atom & SCF & MBPT(2) & CCSD & SCF & MBPT(2)& CCSD \\
\\
\hline
Li$^{+}$ &\,$-$7.237045     &\,$-$0.038520 &\,$-$0.042313&\,$-$7.237174     &\,$-$0.038490&\,$-$0.042284\\
Na$^{+}$ &\,$-$161.885871   &\,$-$0.352072 &\,$-$0.355824&\,$-$161.895637   &\,$-$0.351504&\,$-$0.355258\\
K$^{+}$  &\,$-$601.317915   &\,$-$0.655371 &\,$-$0.669654&\,$-$601.378197   &\,$-$0.654132&\,$-$0.668419\\
Rb$^{+}$ &\,$-$2979.125369  &\,$-$1.603877 &\,$-$1.544953&\,$-$2979.693217  &\,$-$1.600533&\,$-$1.541667\\
Cs$^{+}$ &\,$-$7784.579785  &\,$-$1.746204 &\,$-$1.654556&\,$-$7786.643511  &\,$-$1.740037&\,$-$1.648569\\
Fr$^{+}$ &\,$-$24296.910671 &\,$-$1.635291 &\,$-$1.468312&\,$-$24308.061505 &\,$-$1.634768&\,$-$1.467783\\
\hline
\hline
\end{tabular}
\end{center}
\label{tab1}
\end{table}
\end{landscape}
\begin{landscape}
\begin{table*}[h]
\caption{Bond length(in $\AA$), SCF (${E_{DF}^{0}}$) and correlation energy from MBPT(2) and CCSD of LiX(X=H, F, Cl, Br) and NaY(H, F, Cl).}
\newcommand{\mc}[3]{\multicolumn{#1}{#2}{#3}}
\begin{center}
\begin{tabular}{lcrccrcc}
\hline
\hline
 & & \mc{3}{c}{X2C} & \mc{3}{c}{4C}\\
\cline{3-5} \cline{6-8}\\
Molecule & Bond Length \cite{40}& SCF & MBPT(2) & CCSD & SCF & MBPT(2)& CCSD\\
\\
\hline
LiH &\,1.5957&\,$-$7.987794   &\,$-$0.069223 &\,$-$0.080955&\,$-$7.987928   &\,$-$0.069191&\,$-$0.080923\\
LiF &\,1.5939&\,$-$107.078887 &\,$-$0.400341 &\,$-$0.399171&\,$-$107.084024 &\,$-$0.400097&\,$-$0.398927\\
LiCl&\,2.0207&\,$-$468.469738 &\,$-$0.606016 &\,$-$0.622235&\,$-$468.511539 &\,$-$0.605515&\,$-$0.621738\\
LiBr&\,2.1704&\,$-$2612.134924&\,$-$1.465099 &\,$-$1.404927&\,$-$2612.603993&\,$-$1.463721&\,$-$1.403588\\
NaH &\,1.8874&\,$-$162.602176 &\,$-$0.385821 &\,$-$0.397925&\,$-$162.611952 &\,$-$0.385251&\,$-$0.397357\\
NaF &\,1.9259&\,$-$261.676902 &\,$-$0.694456 &\,$-$0.691459&\,$-$261.691677 &\,$-$0.694011&\,$-$0.691018\\
NaCl&\,2.3608&\,$-$623.077913 &\,$-$0.846128 &\,$-$0.861815&\,$-$623.129354 &\,$-$0.845488&\,$-$0.861179\\
\hline
\hline
\end{tabular}
\end{center}
\label{tab2}
\end{table*}
\end{landscape}
\section{Computational Details}\label{sec3}
The DIRAC program package \cite{32,38} is used to evaluate the required one-body and two-body matrix elements for the correlation calculation.
Both the X2C and four-component calculations are done by using uncontracted finite atomic basis, which
is consists of scalar real gaussian functions.
The small component of the basis is linked with the large component of the basis through
the restricted kinetic balance (RKB) condition.
The RKB condition represents the kinetic energy properly in the non relativistic limit
and avoids the variational collapse \cite{33}.
This is achieved by pre-projecting in scalar basis and unphysical solutions 
are removed by diagonalizing the free particle Hamiltonian.
The RKB condition generates the positronic solution and electronic solution in 1:1 manner.
The DIRAC program package uses Gaussian distribution nuclear model to take care of the 
finite size of the nucleus. The used nuclear parameters are taken as default values from DIRAC package \cite{33a}.
We adopted aug-cc-pCVQZ basis for Li$^+$ \cite{34} and Na$^+$ \cite{35} atom and all the generated orbitals are taken
into consideration for the correlation calculations. Dyall.cv4z \cite{36} basis is opted for K$^+$ and Rb$^+$.
We have neglected the virtual orbitals those energy is more than 500 a.u. for the K$^+$ and Rb$^+$ atom.
The Cs$^+$ and Fr$^+$ are calculated using dyall.cv3z basis \cite{36}. The cutoff used for Cs$^+$ atom is 
1000 a.u. whereas for Fr$^+$ atom, we have taken the orbitals having energy in between $-$25 a.u. 
to 100 a.u. in our correlation calculations.
\begin{table}[t]
\caption{Convergence pattern of electron affinity (in eV) of the C$_{2}$ (R=1.243 \AA{}, Ref \cite{40}) as a function of basis set.}
{
\begin{center}
\begin{tabular}{lccr}
\hline
\hline
Basis & Spinor&  Electron affinity& Expt.\cite{41}\\
\\
\hline
cc-pVDZ &\, 88 &\, 2.6494\\
cc-pCVDZ&\, 104&\, 2.6645\\
aug-cc-pVDZ&\,124&\, 3.1896\\
aug-cc-pCVDZ &\,140&\,3.1949\\
cc-pVTZ&\,152&\,3.1140\\
cc-pCVTZ&\,192&\,3.1285\\
aug-cc-pVTZ&\,216&\,3.3316&\, 3.30$\pm$0.1 \\
cc-pVQZ&\,252&\,3.2743\\
aug-cc-pCVTZ&\,256&\,3.3412\\
cc-pCVQZ&\,316&\,3.2851\\
aug-cc-pCVQZ&\,416&\,3.3840\\
aug-cc-pCVQZ$^{a}$&\,488&\,3.3853\\
\hline
\hline
\end{tabular}
\end{center}
}
$^{a}$ All the virtual orbitals are used for the EOMCC calculation.\\
\label{tab3}
\end{table}
In the molecular calculations of LiF, LiCl, LiBr, we have chosen aug-cc-pCVTZ basis for Li atom \cite{34} and 
cutoff of 100 a.u. for the virtual orbitals.
The calculations of F and Cl are done using aug-cc-pCVQZ \cite{34,35} basis and for Br, dyall.cv4z \cite{37} basis is used. 
In LiH we have chosen aug-cc-pCVQZ basis \cite{34} for Li and aug-cc-pVTZ \cite{34} for the H atom and none of the electrons are frozen for the correlation
calculations. 
Aug-cc-pCVTZ basis is opted for both Na \cite{35} and Cl \cite{34a} in the calculations of NaCl and a cutoff of 100 a.u. is used for the virtual orbitals. 
The single particle orbitals and two-body matrix elements are generated by taking account of C$_{2v}$ symmetry.
Both X2C and four-component calculations of Rb and LiBr are done with the DIRAC14 version and rest of the calculations
are done using DIRAC10. The implemented version of X2C SCF \cite{39} in DIRAC10 is capable of taking up to 
g harmonics but the opted basis for Rb and LiBr require up to h harmonics to express the large component of the wave function. Therefore, these two 
calculations are done using DIRAC14 version.
We have fixed a cutoff of $10^{-12}$ to store the matrix elements for the intermediate diagrams as 
two-body matrix elements contributed negligibly beyond this limit. 
The convergence of $10^{-9}$ is fixed for the solution of SRCC amplitude equations and $10^{-5}$
for the Davidson algorithm in the EOMCC part.
A direct inversion in the iterative subspace (DIIS) of 6 is used in the solution of ground state amplitudes for all the calculations.
\begin{table}[ht]
\caption{Ionization potential values (in eV) of alkali metal atoms.}
\begin{center}
\begin{tabular}{lccr}
\hline
\hline
Atom & X2C & 4C & NIST \cite{42}\\
\hline
Li &\,5.3894&\,5.3895&\,5.3917\\
Na &\,5.1104&\,5.1106&\,5.1391\\
K  &\,4.3419&\,4.3423&\,4.3407\\
Rb &\,4.1750&\,4.1756&\,4.1771\\
Cs &\,3.8861&\,3.8872&\,3.8939\\
Fr &\,4.0579&\,4.0603&\,4.0727\\
\hline
\hline
\end{tabular}
\end{center}
\label{tab4}
\end{table}
The newly implemented relativistic EOMCC code is tested by comparing EA-EOMCC results with the (1,0) sector FSMRCC
code implemented in the DIRAC package as these two theories are supposed to produce identical results for one 
electron attachment process.
The MBPT(2) correlation energy is identical whereas CCSD correlation energy and the EA value are matching upto ten-digit
and eight-digit, respectively.
The test is performed with identical convergence cut off, equal number of DIIS space and without any cutoff in the intermediate diagrams.
We have done the test over a series of atoms and molecules with various basis sets and successful in achieving
similar agreement for all the considered system, independent of the choice of basis set.
\section{Results and discussion}\label{sec4}
We have reserved this section of our manuscript to present numerical results of our calculations and to interpret the outcome of these calculations.
The four-component and exact two component (X2C) EOMCC calculations are performed for all the considered atomic and
molecular systems starting from their closed-shell configuration.\par
%
In Tables \ref{tab1} and \ref{tab2}, we present numerical results of our SCF and correlation
energy from MBPT(2) and CCSD calculation of singly
positive alkali metal atomic systems (Li$^+$, Na$^+$, K$^+$, Rb$^+$, Cs$^+$, Fr$^+$) and
molecular systems (LiX (X=H, F, Cl, Br) and NaY (Y=H, F, Cl)) in their closed-shell configuration.
The bond length of the molecular systems is also compiled in Table \ref{tab2}.
For atomic systems (Table \ref{tab1}), we noticed that the difference between MBPT(2) and CCSD correlation energies
for both X2C and four component calculation keep on increasing as we go down the group. This trend is expected as the
effect of correlation increases as the number of electron in the system increases. 
\begin{landscape}
\begin{table}[ht]
\caption{ Relaive energy difference (in eV) of energy levels of atoms }
\newcommand{\mc}[3]{\multicolumn{#1}{#2}{#3}}
\begin{center}
\begin{tabular}{lcccccccccr}
\hline
\hline
Atom & $^{2}$S & \mc{3}{c}{$^{2}$P$_{1/2}$} & \mc{3}{c}{$^{2}$P$_{3/2}$} & \mc{3}{c}{$^{2}$P$_{3/2}$ - $^{2}$P$_{1/2}$}\\
\cline{3-5} \cline{6-8} \cline{9-11}
× & × & X2C & 4C & NIST \cite{42} & X2C & 4C & NIST \cite{42} & X2C & 4C & NIST \cite{42}\\
\hline
Li & 0.0000 & 1.8494 & 1.8495 & 1.8478 & 1.8495 & 1.8495 & 1.8479 & 0.0001 & 0.0000 & 0.0001\\
Na & 0.0000 & 2.0837 & 2.0840 & 2.1023 & 2.0859 & 2.0862 & 2.1044 & 0.0022 & 0.0022 & 0.0021\\
K  & 0.0000 & 1.6128 & 1.6132 & 1.6100 & 1.6203 & 1.6206 & 1.6171 & 0.0075 & 0.0074 & 0.0071\\
Rb & 0.0000 & 1.5610 & 1.5616 & 1.5596 & 1.5909 & 1.5915 & 1.5890 & 0.0299 & 0.0299 & 0.0294\\
Cs & 0.0000 & 1.3912 & 1.3921 & 1.3859 & 1.4600 & 1.4609 & 1.4546 & 0.0688 & 0.0688 & 0.0687\\
Fr & 0.0000 & 1.5198 & 1.5220 & 1.5172 & 1.7276 & 1.7298 & 1.7264 & 0.2078 & 0.2078 & 0.2092\\
\hline
\hline
\end{tabular}
\end{center}
\label{eng_diff}
\end{table}
\end{landscape}


We have done a series of calculations to understand how the electron affinity 
value changes with the nature of basis set and cutoff in the orbital energies can be used without
losing considerable amount of accuracy as EOMCC calculations for the EA problem are computationally costly. We have chosen C$_2$ as an example 
system for which experimental vertical EA value is reported in the literature. We have started our calculation
with cc-pVDZ, which is a very small basis as it generates only 88 spinor for the beyond SCF calculations using a cutoff of 100 a.u. for the virtual orbitals. 
A few more calculations are also done by improving the nature of the basis functions. The EA value as well as the number of generated spinor in different basis
are tabulated in Table \ref{tab3}.
We have taken 1.243 \AA{} as bond length for the C$_2$ molecule, which is the experimentally reported bond length \cite{40}.
The reported experimental value is 3.30$\pm$0.1 eV \cite{41}, whereas our calculation yields
3.3840 eV for aug-cc-pCVQZ basis with a cutoff of 100 a.u. in the virtual orbital energy.
On the other hand, without using any cut off, the result is 3.3853 eV.
Therefore, a cutoff of 100 a.u. for virtual orbital energies and similar basis set can be used without loosing
much accuracy for all other calculations to achieve a good agreement with the experiment.
It will save enormous computational time without losing a significant amount of accuracy as contribution from the high energy virtual
orbitals is very less in the correlation calculations.
The reported experimental uncertainty is in the first digit after the decimal point. Therefore, it is hard 
to comment on the accuracy of our calculated results. It can be said that
our results are also spanning same range starting from a reasonable basis. 
\begin{table}[t]
\caption{Vertical EA (in eV) values of LiX (X=H, F, Cl, Br) and NaY (Y=H, F, Cl).}
\begin{center}
\begin{tabular}{lccr}
\hline
\hline
Molecule & X2C & 4C & Others \cite{43}\\
\hline
LiH &\,0.2968&\,0.2968&\,0.247\\
LiF &\,0.3550&\,0.3550&\,0.340\\
LiCl&\,0.5526&\,0.5526&\,0.551\\
LiBr&\,0.6148&\,0.6148&\, \\
NaH &\,0.3218&\,0.3217&\,0.319\\
NaF &\,0.4848&\,0.4849&\,0.485\\
NaCl&\,0.6726&\,0.6727&\,0.672\\
\hline
\hline
\end{tabular}
\end{center}
\label{tab5}
\end{table}
In Table \ref{tab4}, we report the calculated ionization potential values of atomic systems using both X2C
and four-component EA-EOMCC method. We have started our calculations from singly positive alkali metal ions
and applied EA-EOMCC method. The negative of the computed values are reported as ionization potential 
values of the open-shell atomic systems.
These computed ionization potential values are compared with the values from the National Institute of Standards and Technology
(NIST) database. A nice agreement with NIST values is achieved for all the considered systems. 
The maximum deviation is obtained for the Na atom, which is also in the accuracy of $\sim$ 0.6\% with the NIST value.
The difference between the X2C with four-component results is in the fifth digit after the decimal for Li atom whereas
the difference is about 0.01 eV for Fr atom. \par

We have calculated electron attachment energy to the $p_{1/2}$ and $p_{3/2}$ orbitals of the 
atomic systems in the X2C and four-component EOMCC framework and presented as a relative energy difference
with respect to the $^2S$ states. These results are compared with the values from NIST database 
and presented in the Table \ref{eng_diff}. The results of our calculations are found to be very accurate 
with respect to the NIST values. It is interesting to note that the energy gap between the $p_{1/2}$ and $p_{3/2}$
is negligible for the Li atom but it keeps on increasing as the system size become havier.
This can be explained by the fact that the effect of relativity increases with the increase in atomic number
and thus, the spin-orbit coupling plays a significant role in heavier systems.
\par
In Table \ref{tab5}, we present the results of our calculations of vertical EA of molecular systems using 
both X2C EA-EOMCC and four-component EA-EOMCC method. Further, we have compared our result with the theoretical results
calculated by Gutsev {\it et al} \cite{43}. They also employed EA-EOMCC method for correlation treatment to
calculate the vertical EA values of the molecular systems. In their calculation, Gutsev {\it et al} misses the
effect of relativity, which is included in our calculation in its four-component formalism.
We have achieved a nice agreement for the atomic results and also for the vertical EA 
value of C$_2$ molecule. Therefore, it can be said that our calculated results for the 
molecular systems are also quite accurate though there is no reliable experimental data
or any other values calculated using any variant of relativistic coupled cluster theory
to compare with. However, the accuracy of the molecular calculations will not be that much accurate
as compared to the atomic results. The reason behind this is due to the possibility of structural 
change on attachment of an extra electron to the neutral molecule depending on the polarity 
of the molecule. \par
\section{Conclusion}\label{sec5}
The relativistic EOMCC method for the electron attachment problem applicable to both atomic and molecular systems is
successfully implemented. To test the performance of the EA-EOMCC method, we applied to calculate ionization potential values of 
alkali metal atoms starting from closed-shell configuration. We have compared our calculated ionization 
potential values with the values from NIST database. We are successful in achieving less than 1\% agreement with the
NIST values. We have also presented molecular EA values of LiX (X=H, F, Cl, Br) and NaY (Y=H, F, Cl) using our 
relativistic EOMCC methods.
\section*{Acknowledgements}
Authors acknowledge a grant from CSIR XIIth Five Year Plan project on Multi-Scale Simulations of 
Material (MSM) and the resources of the Center of Excellence in Scientific Computing at CSIR-NCL.
H. P. acknowledges the  Council of Scientific and Industrial Research (CSIR) for fellowship.
S. S. acknowledges the  Council of Scientific and Industrial Research (CSIR) for Shyama Prasad Mukherjee Fellowship.
S. P. acknowledges funding from J. C. Bose Fellowship grant of Department of Science and Technology (India).




\section*{References}


\end{document}